# Role of oxygen interstitials in $Zn_{1-x}Ga_xO$ for faster response to UV light

Prashant Kumar Mishra[1], Tulika Srivastava[2], Saniya Ayaz[3], Ramraj Meena[4], Sajal Biring[5], Somaditya Sen[6,*]

**Abstract**

ZnO doped with Gallium ($Ga^{3+}$) demonstrates better crystalline nature and conductivity increases. Latent defect states are suppressed. However, due to the larger charge of $Ga^{3+}$ oxygen interstitials are generated which control the sensing speed. The conductance increases as a consequence of reduced defect states, especially the oxygen vacancies. The photocurrent increases with Galium incorporation, but a more intense increase in the current reduces the sensitivity.

**Keywords**: *Gallium, ZnO, Defects, UV sensitivity, Response, Recovery*

Zinc oxide (ZnO) is a well-known non-toxic II-VI semiconductor [1] oxide with bandgap, $E_g \sim$ 3.37eV [2], which can be modified by proper substitution [3]. Substitution can improve optoelectronic [4], photoconductive [4, 5] and sensing properties [6] of ZnO. Transport properties of ZnO are modified in UV light due to surface defects [6,7], grain size [5-7] and electron adsorption/desorption properties [5, 6, 7]. This makes ZnO extremely competent in UV sensing and response. Surface defects may comprise of oxygen related defects such as oxygen vacancies (Vo) [8] and oxygen interstitials ($O_i$) [7, 8] or zinc related defects such as zinc vacancies ($V_{Zn}$) [10] and zinc interstitials ($Zn_i$), etc. [8, 9].

Higher valent dopant/substituent ions have a tendency of drawing more oxygen to the lattice [6-8]. This results in reduction of latent defects of ZnO, such as Vo [8] and introduction of $O_i$ [7, 9]. Other latent defects such as $V_{Zn}$ [6, 7, 8] and $Zn_i$. Doping of B [11], In [12], Al [13] and Ga [13-15] are found in literature, reporting sensing [16] and other optoelectronic properties [15]. Ga plays an important role because of its lower reactivity, greater resistance to oxidation, and higher electro-negativity. ZnO lattice deformation is small even with high Ga concentration [11-15]. The conductivity increases due to extra charge of $Ga^{3+}$ [15-17], making it suitable as a transparent conductive oxide (TCO) materials . Different synthesis methods have been reported via different synthesis routes [11, 18]. Sol-gel is one of them [17].

A complete understanding of the dynamics of defect formation in the bandgap due to $Ga^{3+}$ incorporation and its role on UV-sensing properties is studied in this work. A single phase homogeneous substitution is ensure by adopting a low-cost yet certain synthesis route using a sol-gel (Pechini) method [18]. Gallium-substituted ZnO nanoparticles $Zn_{(1-x)}Ga_xO$ for x= 0 (ZG0), 0.01 (ZG1) and 0.03 (ZG3) have been synthesized. Defects or electron trap states, present between conduction band (CB) and valence band (VB) are evaluated using optoelectronic techniques. The defects have been individually discussed and a correlation been proposed in this work.

Zn-solution was prepared by dissolving ZnO powders (Alfa Aesar, purity 99.9%) in $HNO_3$ (Alfa Aesar, purity 99.9%). Ga-solution was prepared by dissolving using Gallium nitrate in DI water. A homogeneous Zn/Ga solution was prepared by mixing these two solutions and stirring for ~1hr. In a separate beaker, citric acid and glycerol were mixed in D.I. water and stirred for 1hr to form polymers and to be used as gelling agent. The Zn/Ga solution was added to this polymeric solution and continuously stirred and heated at ~70° C for ~4hrs. During this process, the Zn/Ga ions get attached homogeneously to the polymer solution. Upon prolonged heating gel was formed. The gels were burnt on hot plates in ambient conditions. The burnt gels were denitrified and decarbonized by heating in muffle furnaces at ~450°C for ~6hrs to produce yellowish white nanopowders of the desired composition. Further, they were annealed at 600°C for 2 h.

Wurtzite structures of all the ZG samples are confirmed without any secondary phase [Figure 1 (a)] from X-ray diffraction (XRD) studies using Bruker D2-Phaser diffractometer, with a Cu Kα (wavelength = 1.5406Å) X-ray source. Rietveld refinement of samples with wurtzite P63mc space group was performed using GSAS software [Figure 1(a)-inset]. Lattice parameters changes insignificantly, leading to nominal increment in c/a ratio [Figure 1(b,c)].

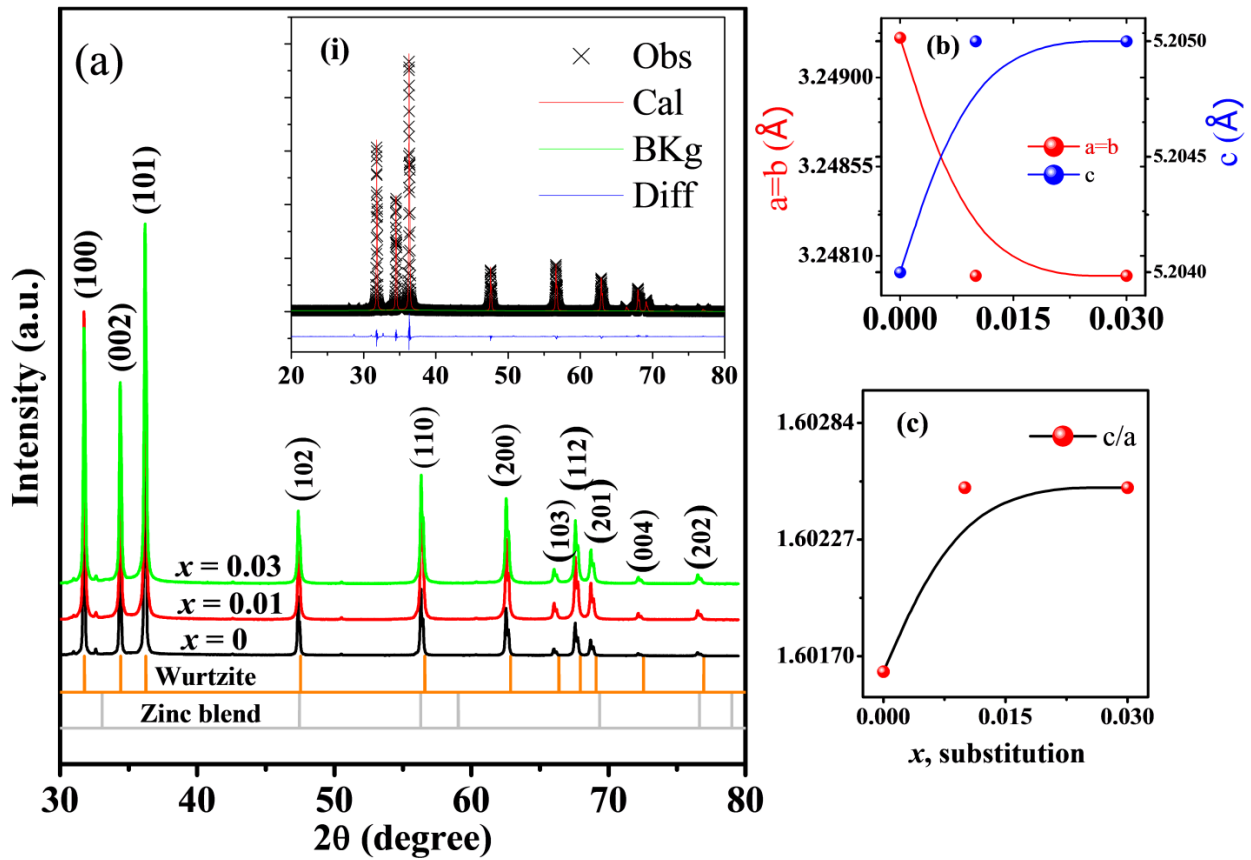

*Figure 1(a). XRD pattern of samples ZG0, ZG1, ZG3 (-inset Rietveld of sample ZG0) (b) insignificant variation of lattice parameters and (c) nominal increment in c/a ratio*

Optical reflectance was collected using a UV-VIS-NIR Shimadzu (UV-3600) Spectrophotometer. $E_g$ was estimated [Figure 2(a,b,c)] using *Tauc plot* method [8, 18]: $\alpha \cdot h\nu = A \cdot (h\nu - E_g)^n$, where, $\alpha$ is absorption coefficient, $A$ is a constant, $h\nu$ is the energy of the incident light and $n$ is number ($n=1/2$ for direct and $n=2$ for indirect bandgap). Urbach energy ($E_U$), an estimate of lattice disorder [18], was calculated from the slopes of $\ln\alpha$ versus $h\nu$ graph where, $\alpha = \alpha_0 \exp(h\nu/E_U)$, where, $\alpha_0$ is a constant [Figure 2(d, e, f)]. On Ga doping, $E_g$ increases nominally, from ~3.21 eV (ZG0) to 3.24 eV (ZG3), while $E_U$ decreases from ~93 meV (ZG0) to 84 meV (ZG3) [Figure 2(g)]. Hence, it seems, Ga doping helps to stabilize the structure by reducing latent defects in the lattice.

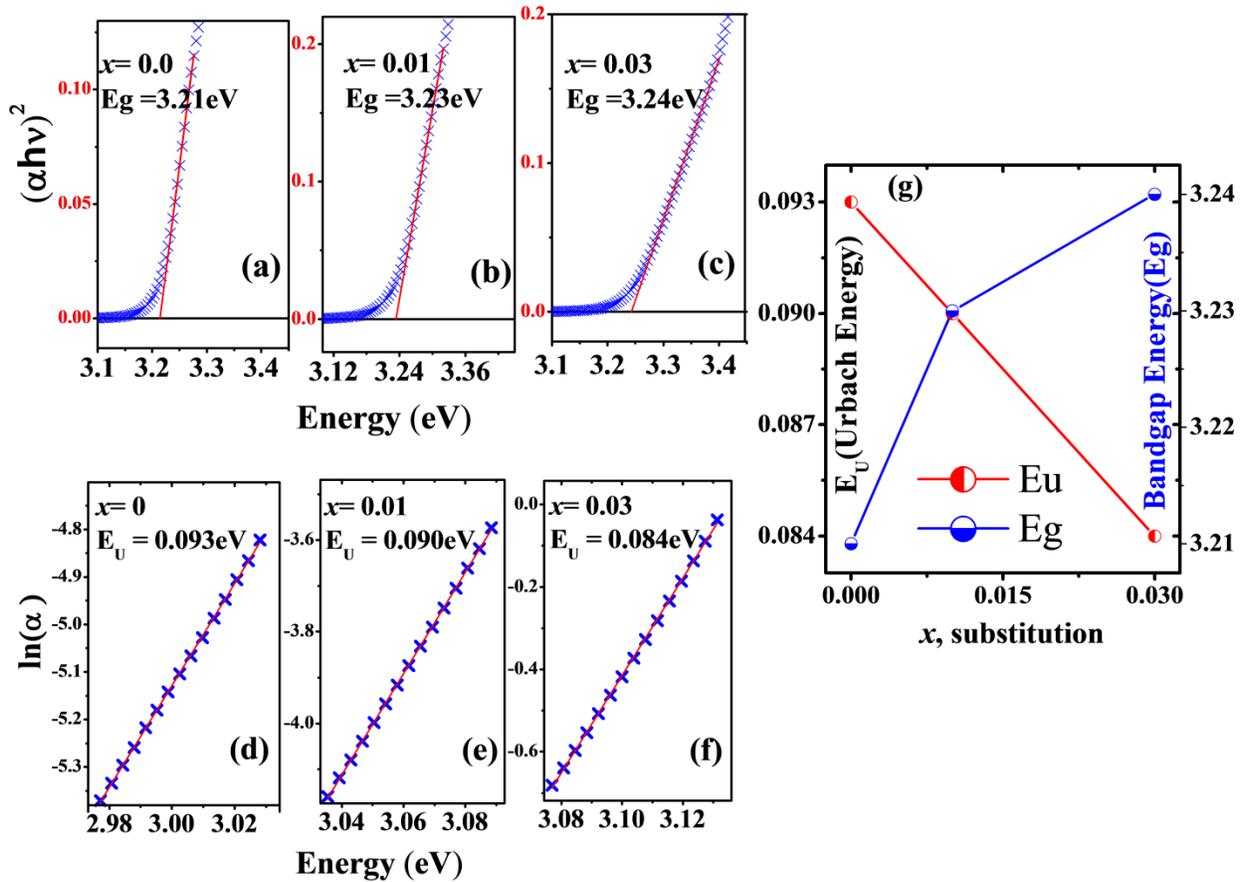

*Figure 2(a). Fitted linear portion between $(\alpha h\nu)^2$ vs $h\nu$ for bandgap estimation (b) linear fit of $\ln(\alpha)$ vs $h\nu$ for $E_U$ (c) $E_g$ and $E_U$ vs x (substitution)*

Defect states in the bandgap of all samples were investigated at room temperature from the photoluminescence spectra (PL) using a Dongwoo Optron DM 500i Spectrometer. The PL spectra comprises of NBE (near band emission, i.e. UV part) and DLE (deep level emission, i.e. visible color emission) [Figure 3(a)]. The DLE and NBE spectra were deconvoluted into various emission peaks in different color regions [Figure 3(b,c,d)], e.g., UV (>3.1 eV), violet (~3–3.1 eV), blue (2.50–2.75 eV), green (2.17–2.50 eV), yellow (2.10–2.17 eV) and orange-red (2.1 eV) [8]. These emissions correspond to various defects, e.g., $Zn_i$ (violet), $V_O$ (green), $O_i$ (yellow, orange-red), and $V_{Zn}$ (blue) [8,19], located below the CB at ~0.22 eV, ~2.5 eV, ~2.28 eV, and ~2.85 eV, respectively.

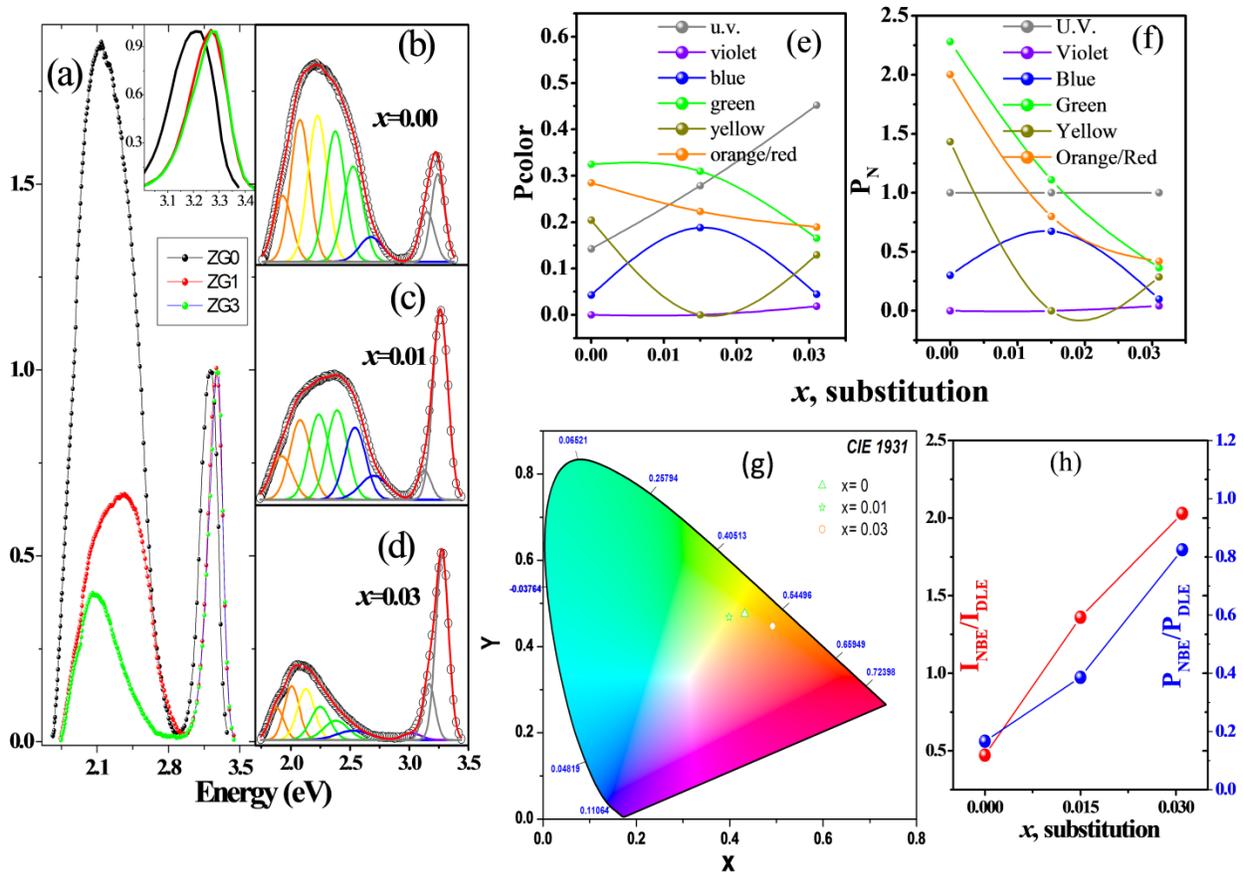

*Figure 3(a). DLE and NBE region for samples ZGO, ZG1, ZG3 respectively (b, c, d) De-convoluted DLE and NBE for samples (e) Pcolor vs x (f) $P_N$ vs x (g) CIE representation of samples (h) $I_{NBE}/I_{DLE}$ and $P_{NBE}/P_{DLE}$ vs x*

Intensity of DLE region as a whole decreases with Ga doping indicating reduction in defect states [Figure 3(a)]. The gradual shift of the NBE to higher energy [Figure 3(a) inset] confirms the changes in $E_g$ in agreement with UV-vis spectroscopy results. Area calculation of each peak helps understand color contribution [8,19,20]. The proportion of each color in PL spectrum is estimated as $P_{color}=A_{color}/A_{total}$ [Figure 3(e)], such that $A_{total}= \Sigma A_{color}$ and, $\Sigma P_{color}=1$. Individual color contributions, $P_{color}$, have been normalized with respect to $P_{UV}$, to yield $P_N$ [Figure 3(f)] [8], assuming that the UV contribution is not deviated for all samples which is a contribution of a CB→VB transition. It is observed that $P_N$ decrease for green and orange-red colors. The decrease in green color intensity hints at a reduction in latent $V_O$. Notably, latent $O_i$ (orange-red emission) decreases at par Vo. Hence, the latent Oi present may be filling up the latent Vo present in pure ZnO. $Zn_i$ (presence of violet emission) is present only in ZG3. $V_{Zn}$ (blue emission) are present in ZG0 and increases in ZG1 samples, but nearly vanishes in ZG3. Yellow emission too shows such a fluctuating nature and decreases drastically for ZG1 but thereafter increases for ZG3. The origin of this type of $O_i$ may be different than the $O_i$ due to orange-red emission. The inverse trends between $V_{Zn}$ and $O_i$ tempts one to believe that these $O_i$ were actually occupying a position adjacent to a Zn-site in the pure ZnO, which by the substitution of a smaller $Ga^{3+}$ ion (0.61A) has been rearranged, creating $V_{Zn}$. Beyond ZG1, the extra charges of $Ga^{3+}$ ions may attract excess O ions, which will result in $O_i$, in the vicinity of a comparably more attractive $Ga^{3+}$, thereby filling in the $V_{Zn}$ [Figure 3(f)].

A CIE graph (1931 standard) was calculated using OSA software. Chromaticity coordinates obtained were as follows: ZG0 (0.39, 0.47), ZG1 (0.43, 0.46) and ZG3 (0.49, 0.44) [Figure 3(g)]. A drastic coordinate shifting toward orange-red emission is observed for ZG1 and further for ZG3. Shifting of coordinates is probably due to a proportionate reduction of Vo and Oi, and a simultaneous extra addition of newly created $O_i$ due to the extra charge of the substituent $Ga^{3+}$ [6,8]. Therefore, it can be inferred that all sorts of latent defects reduce with Ga doping while $Ga^{3+}$ associated $O_i$ increases for ZG3. Both $I_{NBE}/I_{DLE}$ (intensity ratio) and $P_{NBE}/P_{DLE}$ (color ratio) ratios increase with $Ga^{3+}$ substitution which indicate latent defects suppression [Figure 3(h)]. This is agreement with UV-vis studies.

Pellets of 1 mm thickness and 10 mm diameter were uni-axially pressed under 3 Tons pressure. Two electrodes were prepared at a distance of 1 mm on the flatter surface of pellets using silver paste. UV light sensing was investigated by measuring the changes in current at constant voltage using a home-made set-up and a Keithley (2401) electrometer. The samples were kept inside a dark box without any stray light. A steady UV light source of 390 nm wavelength was focused in between the electrodes from a hole at the top of the black box. UV light was switched ON (for 7.5min) and OFF (for 7.5min) for four cycles and dynamic changes in current with respect to time were measured [Figure 4(a)]. The current increased bi-exponentially with UV illumination to a maximum current [Figure 4(b)], $I_{UV\_max}$ and decreased similarly after light was turned OFF to a minimum current in darkness, $I_{dark\_min}$ [Figure 4(c)]. Notably, $I_{dark\_min}$ continually increases with Gallium doping [Figure 4(e)], implying increase of the conductance of the Ga-doped samples. The sensing photocurrent, $\Delta I$, is the difference between $I_{UV\_max}$ and $I_{dark\_min}$; i.e. $\Delta I = I_{UV\_max} - I_{dark\_min}$ also increases but the increase in $I_{dark\_min}$ is comparably more [Figure 4(e)]. Sensitivity, S, of samples may be defined as $S = (\Delta I/I_{dark\_min})*100$. As DI increases in lesser rate than $I_{dark\_min}$, S decreases with increase in $Ga^{3+}$ doping [Figure 4(e)].

The fastness of UV-sensing can be estimated by normalizing the change of current for the three systems. For this a plot of $(I-I_{dark\_min})/\Delta I$ has been shown in Figure 4 (c). The pure ZG0 system seems to be the fastest while ZG1 seems to be the slowest amongst the three samples. To be noted that for ZG1, $O_i$ defects are the least while $V_{Zn}$ are maximum. This similarity tempts one to assume a direct correlation between availability of Oi defects, or absence of $V_{Zn}$ and the response to UV light.

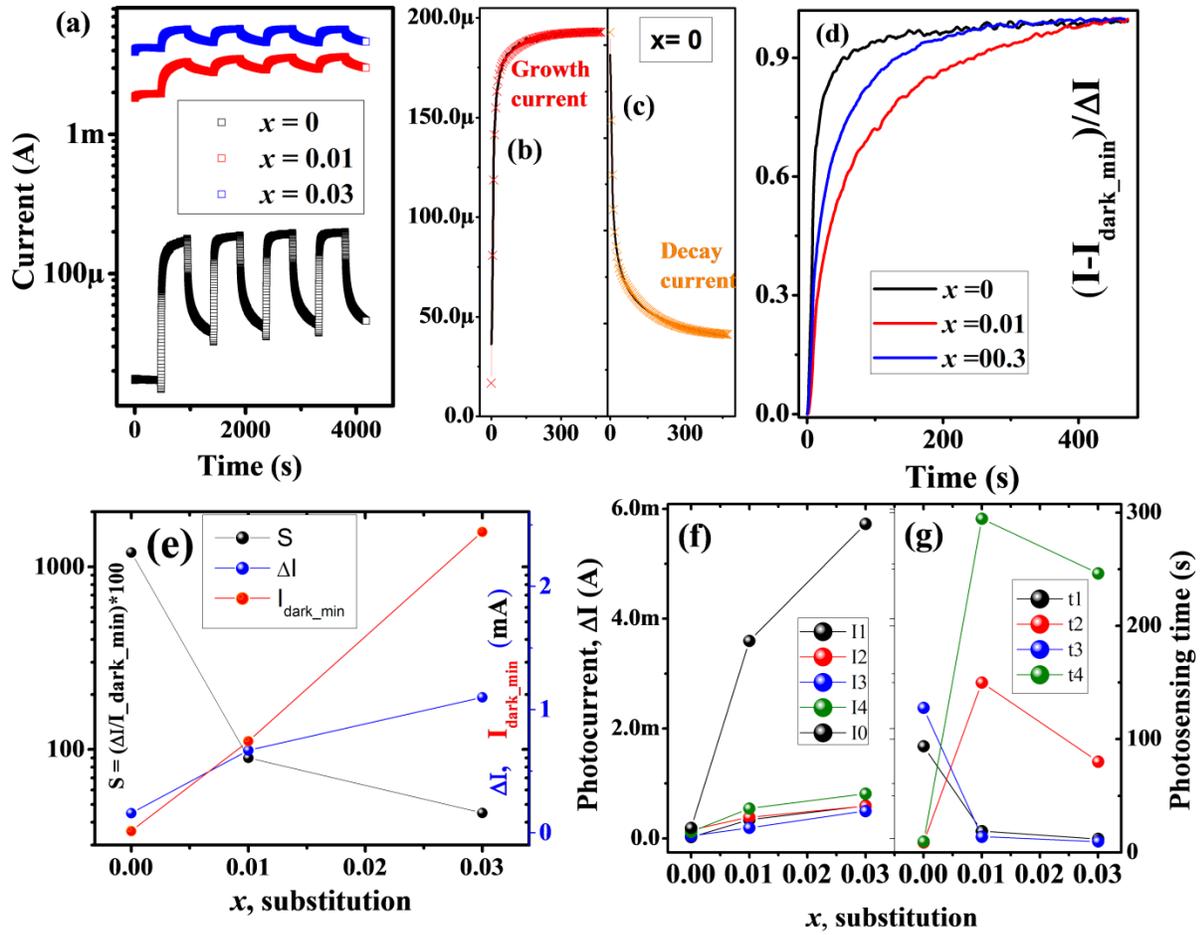

*Figure 4(a) dynamic change in current with x (b, c) Bi-exponential fitting of grow and decay current vs time ZG0 (d) $(I-I_{dark\_min})/\Delta I$ with time (e) S, $\Delta I$ and $I_{dark\_min}$ vs x (f) $\Delta I$ and I0 with x after bi-exponential fitting of two parts grow and decay (g) UV sensing time with substitution*

A bi-exponential formula for the current, $I_g = I_{g0} + I_1 \exp(-t/t_1) + I_2 \exp(-t/t_2)$ was fitted to both growth of current, where $I_{g0}$ is the current offset; $I_1$, $I_2$ are positive constants; $t_1$ and $t_2$ are the fast and slow time constants of response [21]. Similarly, $I_d = I_{d0} + I_3 \exp(-t/t_3) + I_4 \exp(-t/t_4)$ was used for current decay. It was observed that, while $t_1$ decreases with substitution from $t_1 = 93.77s$ (Z0) to 11.90s (ZG3), $t_2$ increases from $t_2 = 8.83s$ (Z0) to 246.34s (ZG3). Therefore, while the faster process becomes ~8 times faster, the slower process becomes ~28 times slower with Ga-doping. Similarly, while $t_3$ decreases from 127.76s (Z0) to 9.61s (ZG3), $t_4$ increases from 9.63s (Z0) to 294.60s (ZG3). Therefore, while the faster decay process becomes ~13 times faster, the slower process becomes ~30 times slower with Ga-doping. It is noteworthy that $t_2$ and $t_4$ are maxima for

ZG1 whereas $t_1$ and $t_3$ decreases with Gallium doping. [Figure 4(g)]. This behavior of variance of $t_1$ and $t_3$ matches that of $O_i$ defects, whereas the nature of other defects matches with that of $t_2$ and $t_4$.

The currents associated with these two processes increases continually with increasing doping irrespective of the process, and are comparable to each other [Figure 4(f)]. Hence, these two processes are equally dominant. The sensing being faster in pure ZnO is hence more appropriately dependent on the faster processes, and most probably due to the $O_i$ defects as in ZG1 the sensing process is the least with $O_i$ being the least as well.

Oxygen-related surface defects enhance photo-sensing [21-22]. Adsorption /desorption of electron at the surface changes with UV illumination. Oxygen molecules capture free electrons from the surface of pellets and thereby get adsorbed [$O_2$ (g) +$e^-$→$O_2^-$(ad)]. In the presence of UV light, electron-hole pairs are generated [hν→$e^-$+ $h^+$] on the surface of the pellets [6]. These holes react with adsorbed oxygen, liberating oxygen from the surface [h+$O_2^-$(ad) →$O_2$ (g)].With reduction in $V_O$ and $O_i$, sensitivity is expected to decrease [6, 21, 22]. However, from the formula of sensitivity it seems that it is strongly dependent on $I_{dark\_min}$ which is comparable to $I_0$ in this case. Faster response depends on the Oi density. However, sensitivity is highly dominated by $I_{dark\_min}$. In spite of sensitivity being lowered with Ga-doping, the photocurrent, ΔI does increase with substitution.

In this work it was observed with certainty that the relative proportion of Oi decrease with nominal introduction of Ga but with further addition increases. As $Ga^{3+}$(IV) (0.61A) is smaller than $Zn^{2+}$(IV) (0.74A), the native Vo are filled up by native and added Oi. However, due to the extra charge of $Ga^{3+}$ more O is incorporated and as a result, number of $O_i$ increase. The speed of response is highly dependent on this factor. The availability of extra charge in $Ga^{3+}$ helps improve the transport properties of the materials.

## Conclusions:

Single phase wurtzite $Zn_{(1-x)}Ga_xO$ samples reveal insignificant changes in lattice parameters with nominal increase in $E_g$, on Ga substitution. Urbach energy $E_U$ decreases with Ga substitution indicating loss of structural disorder. Ga doping helps to stabilize the structure by reducing defects in the lattice. From PL study it was revealed that all sorts of defects reduce with Ga doping. However as the latent defects decrease drastically, $O_i$, (yellow) associated with $Ga^{3+}$ decreases for ZG1 compared to ZG0 but thereafter again increases for ZG3. Photoresponse is faster in pure Z0 system while ZG1 seems to be the slowest amongst the three samples. To be noted that for the ZG1 sample Oi defects are the least while $V_{Zn}$ are maximized. This similarity tempts one to assume a direct correlation between availability of Oi defects, or absence of $V_{Zn}$ and the response to UV light. Sensitivity decreases with increase in $Ga^{3+}$ doping. $I_{dark\_min}$ increases with doping. However, proportionately, the photocurrent ΔI also increases with doping. Hence, the system is extremely dependent on the Oi density for faster response, while sensitivity dependents on the dark current.

## Acknowledgement

The authors are thankful to Dr. Vipul Singh for providing PL spectroscopy and Dr. L.P. Purohit for providing UV-vis spectroscopy. The authors are also thankful to DST for providing Inspire Fellowship.